\documentclass[aps,pre,reprint,groupedaddress]{revtex4-1}

\usepackage{graphicx}
\usepackage[caption=false]{subfig}

\usepackage{amsmath}

\usepackage{hyperref}
\usepackage{color}

\begin{document}

	\title{Classical nucleation theory in the phase-field crystal model}
	
	
	\author{Paul Jreidini}
	%\thanks{}
	\affiliation{Department of Physics, and Centre for the Physics of Materials, McGill University}
	
	\author{Gabriel Kocher}
	\affiliation{Department of Physics, and Centre for the Physics of Materials, McGill University}

	\author{Nikolas Provatas}
	\affiliation{Department of Physics, and Centre for the Physics of Materials, McGill University}
	
	
	\date{\today}
	
	\begin{abstract}		
		A full understanding of polycrystalline materials requires studying the process of nucleation, a thermally activated phase transition that typically occurs at atomistic scales. The numerical modeling of this process is problematic for traditional numerical techniques: commonly used phase-field methods' resolution does not extend to the atomic scales at which nucleation takes places, while atomistic methods such as molecular dynamics are incapable of scaling to the mesoscale regime where late-stage growth and structure formation takes place following earlier nucleation. Consequently, it is of interest to examine nucleation in the more recently proposed phase-field crystal (PFC) model, which attempts to bridge the atomic and mesoscale regimes in microstructure  simulations. In this work, we numerically calculate homogeneous liquid-to-solid nucleation rates and incubation times in the simplest version of the PFC model, for various parameter choices. We show that the model naturally exhibits qualitative agreement with the predictions of classical nucleation theory (CNT) despite a lack of some explicit atomistic features presumed in CNT. We also examine the early appearance of lattice structure in nucleating grains, finding disagreement with some basic assumptions of CNT. We then argue that a quantitatively correct nucleation theory for the PFC model would require extending CNT to a multi-variable theory.
	\end{abstract}
	
	%\keywords{}
	
	\maketitle
	
	%\section{}
	%\subsection{}
	%\subsubsection{}
	
	%%%%%%%%%%%%%%%%%%%%%%%%%%%%%%%%%%%%%%%%%%%%%%%%%%%%%%%%%%%%%%%%%%%%%%%%%%%%%%%%%%%%
	
	\section{Introduction\label{sec:intro}}
	
	The fields of materials science and engineering are built on a fundamental understanding of non-equilibrium phase transitions, which govern microstructure evolution, and hence the properties, of most materials. The rapid pace of technological progress requires materials with ever more demanding specifications on microstructure. This in turns necessitates improved phase transition models to guide the design of novel materials. The advent of plentiful and inexpensive computing power in the past few decades has greatly benefited this endeavor by allowing the numerical study of phase transition problems that prove intractable otherwise.

	In particular, great effort goes into the solidification modeling of polycrystalline materials, which are comprised of numerous microscopic interlocked crystal grains (aka. crystallites) of differing sizes, shapes, orientations, and compositions. These materials include most metals and alloys, as well as some ceramics and polymers, and even a few biological microstructures \cite{jin03}. The morphological and chemical properties of the constituent crystal grains have a direct effect on characteristics of the macroscopic material \cite{granasy_phasefieldnuc,boettinger00}, hence the interest in modeling their formation and evolution. The first step in the grains' formation is the process of nucleation, wherein thermal fluctuations in a progenitor phase stochastically create stable nuclei of a new phase which then proceed to grow. The main difficulty in modeling this process is due to the large range of scales involved: though final grains range in size from a few nanometers to above millimeters depending on the material, the initial nuclei form on atomic lengthscales. Further, some systems are known to exhibit nucleation events simultaneously with large-scale structure evolution, such as in rapidly cooling metal pools formed by laser-beam welding \cite{david03,boettinger00}, and during columnar to equiaxed transition \cite{kurz01}. While the growth and late evolution of grains are relatively well understood \cite{granasy_02}, there remain open questions concerning how to efficiently and accurately model the initial nucleation stage of solidification  without impairing the modeling of the much larger scale evolution. 
	
	Various numerical techniques for simulating phase transition processes, including nucleation, have been developed since the 1960s, each best suited to specific types of problems. Among these, traditional phase-field methods \cite{hohenberg77,singer08,chaikin_condensed,provatas_PFC} are some of the more widely used in studying microstructure formation. They are well-adapted for simulations on micrometer to millimeter length scales, and on diffusive time scales. These methods consist of multi-field descriptions of phases separated by diffuse interfaces, effectively greatly refined versions of Landau's order parameter theory of phase transitions. The fields are spatially continuous, constant within bulk regions, and vary smoothly but rapidly across interfaces. Typically, the fields used can represent order, concentration, and temperature. Phase-field methods have been used to simulate assorted material phenomena that include eutectic (multi-phase solid mixture) system growth \cite{grant94,Fol05,Nestler15}, dendritic microstructure evolution \cite{Kar97,provatas98,kim99,karma01_2,Ech04,Gre04,Ram04,Ech09,Morteza10}, fracture growth \cite{karma01}, and structure changes in irradiated materials \cite{li17}. However, they face difficulty in modeling physical processes that involve atomistic lengthscales. For example, phenomena that occur on the scale of the width of phase interfaces need special care \cite{elder01,Ech04,plapp11}, as  phase-field models approximate interfaces to be much more diffuse than the typical dozen atom-widths found in real materials, for reasons of computational efficiency. Moreover, effects related to crystalline lattice structure, such as orientation and elastic deformation, do not appear naturally in such basic models, instead requiring more coupled fields to be added \cite{oforiopoku2010,steinbach06,granasy_phasefieldnuc,plapp11} and thus increasing computational and mathematical complexity. As nucleation is a fundamentally atomistic process, phase-field models are incapable of modeling its physics even qualitatively: the processes leading to formation of physically-accurate nuclei can not be resolved at these models' scales. Workarounds to this limitation involve either using unrealistically large thermal fluctuations that force ``nucleation" of effective solid domains at the desired time and length scales, or artificially adding already-formed nuclei to the system according to assumed statistics for the material \cite{granasy_phasefieldnuc,granasy_02,plapp11,Montiel12}.

	On the other end of the scale spectrum for modeling techniques, atomistic models, such as the Molecular Dynamics (MD) methods \cite{alder59,rapaport_md}, are capable of simulating phenomena difficult to access with phase-field methods. These include amorphous solidification of metals \cite{ozgen04,tian08}, properties of atomically-rough interfaces \cite{hoyt01}. Notably, MD methods have had recent success in simulating nucleation and nanoscale grain growth in metals \cite{shibuta15}. These methods typically involve tracking the individual positions and interaction potentials of all the atoms in a system, calculating the dynamics of the resulting N-body problem by numerical integration of Newton's equations of motion. However, the large number of atoms tracked by these methods limits the time and length  scales that can reasonably be simulated with current computational speeds to nanometers and nanoseconds respectively. This prevents the scaling of atomistic models for the study of mesoscale structure dynamics, including those that would result from or occur concurrently with nucleation-initiated phase transitions.
	
	Recently, the phase-field crystal (PFC) methodology \cite{elder02,elder04,provatas07,emmerich12,humadi13,provatas_PFC} has emerged as a modified phase-field method that aims to bridge the gap between atomistic models such as MD and mesoscale models such as traditional phase-field methods. Similar to traditional phase-field models, the PFC model represents material with a continuous spatial field. However, this field is now periodic in bulk regions, instead of constant. The periodic field acts as an atomic density field, with its peaks denoting the most likely position of the crystal lattice's atoms. Its amplitude represents a phase's order, with the field having zero amplitude in liquid phases and nonzero in solid phases. In contrast to traditional phase-field methods, the PFC method does not have difficulty in describing aspects of the crystal lattice structure, including different grain orientations, grain boundary dynamics \cite{elder08}, lattice defects, and elastoplasticity \cite{elder02,Berry12,berry14}. Further, unlike in `true' atomistic models, atomic movement on vibrational timescales in the PFC model is effectively averaged out, leaving only movement on diffusive timescales. It has been shown \cite{grant08} that applying coarse-graining in time on atomistic simulation methods such as MD methods recovers similar results as the PFC model. Compared to MD methods, PFC is computationally more efficient due to the lack of tracking of individual atoms. This allows studying phenomena appearing at longer time and length scales than MD is reasonably able to simulate \cite{dantzig12}.
	
	For the reasons stated above, the PFC model can be useful as an intermediate model between atomistic simulations and the more coarse-grained traditional phase-field methods that do not retain atomic scale details. It is thus of interest to examine whether this model can be used to study the process of nucleation without a loss of efficiency or accuracy. Nucleation is known to occur naturally in the PFC model through the inclusion of thermal fluctuations obeying the fluctuation-dissipation theorem, with the resulting nuclei consisting of few `atoms' as would be expected in a physical system. More specifically, Granasy, Tegze, Toth, and Pusztai have studied numerous aspects of nucleation in the PFC model, including nucleation energy barriers, possible amorphous precursor phases, and heteroepitaxy \cite{toth10,toth11}. However, it is yet unclear whether the PFC model can reproduce the time-dependent statistics of the nucleation process predicted by classical nucleation theory (CNT), such as the scaling of nucleation rate and incubation time with temperature. Moreover, the morphology of forming nuclei in the PFC model, as well as their evolution pathway to stable crystal grains, is still poorly understood. The purpose of this work is thus to numerically study nucleation rate and incubation times in the most basic two-dimensional version of the PFC model and to compare the results with the predictions  of classical nucleation theory. We also examine the morphology of stable nuclei in the PFC  model, as well as their early-time behavior.  
	
	The remainder of this work is structured as follows. Section \ref{sec:pfc} briefly presents the simplest PFC model's free energy functional and time-evolution partial differential equation (PDE). Section \ref{sec:cnt} introduces the concepts of classical nucleation theory used to obtain the expected scaling of nucleation rate and incubation time with temperature, and details the method used to compare these scaling relations to those predicted by the PFC model. Section \ref{sec:res} presents the results of our numerical investigation on nucleation rates, incubation times, and nuclei morphology in the PFC model. This is followed by our concluding summary and thoughts in section \ref{sec:con}.
	
	%%%%%%%%%%%%%%%%%%%%%%%%%%%%%%%%%%%%%%%%%%%%%%%%%%%%%%%%%%%%%%%%%%%%%%%%%%%%%%%%%%%%
	
	\section{The phase-field crystal model\label{sec:pfc}}
	
	\subsection{Dimensionless free energy functional\label{ssec:f_func}}
	
	We derive the PFC model's free energy functional from classical density functional theory (CDFT) of solidification as proposed by Ramakrishnan and Yussouff \cite{ramakrishnan79}, and later obtain the PFC model's time-evolution PDE from this functional. The derivation presented below is for a two-dimensional system consisting of a single atomic species capable of existing in a liquid phase and a solid phase, where the solid phase exhibits a triangular lattice structure. This derivation can be extended to system in three dimensions, with more than one atomic species, and with more complicated lattice structures \cite{provatas07,provatas_PFC,greenwood10,Wu10b,greenwood11_2}.
	
	The CDFT provides as a starting point a Helmholtz free energy functional $\mathcal{F}[\rho]$ where $\rho(\vec{r})$ is the local number density of atoms in the system at position $\vec{r}$. Ramakrishnan and Yussouff obtain this free energy by expanding the full energy functional close to a reference liquid state in coexistence with a solid. Taking the reference liquid's density to be $\rho_o$ and defining $\delta\rho(\vec{r})=\rho(\vec{r})-\rho_o$, they show that
	\begin{multline}\label{eq:ramakF}
	\frac{\mathcal{F}}{k_B T}= \int \left\{ \rho \ln\left(\frac{\rho}{\rho_o}\right)-\delta\rho \right\}d\vec{r} \\- \sum_{n=2}^{\infty} \frac{1}{n!} \int \prod_{i=1}^{n} d\vec{r}_i \delta\rho(\vec{r}_i)C_n(\vec{r}_1,\vec{r}_2,\vec{r}_3, ... ,\vec{r}_n)
	\end{multline}
	where the integrals are over the volume of the system. $T$ is the temperature of the system, assumed to be constant through space, and $k_B$ is the Boltzmann constant. The functions $C_n$ are the $n$-point direct correlation functions of the liquid phase. In this derivation, we truncate the integral series up to the two-point correlation function $C_2$, simplified to $C_2(\vec{r}_1,\vec{r}_2)=C(|\vec{r}_1-\vec{r}_2|)$ due to the liquid phase being isotropic, where we have dropped the subscript for convenience. In general, the Fourier transform $\hat{C}(k)$ of the two-point correlation function of a liquid formed of atoms that interact by the Lennard-Jones potential exhibits a rapidly decaying periodic shape \cite{mandel70}, due to the lack of long-range order. We fit a polynomial function in Fourier space that matches only the first peak of the full function, approximating
	\begin{equation}\label{eq:correlationFourier}
	\hat{C}(k) \approx -\hat{C}_0+\hat{C}_2 k^2 - \hat{C}_4 k^4
	\end{equation}
	where $\hat{C}_0$, $\hat{C}_2$, and $\hat{C}_4$ are positive constants chosen so that the peaks match in position and height. The position of the peak in Fourier space determines the fundamental wavelength-scale of the resulting crystalline solid's reciprocal lattice. As there is only a single wavelength-scale, this approximate one-peak correlation function leads to a triangular lattice structure, the simplest two-dimensional Bravais lattice. A different choice for the correlation function can lead to more complex lattice symmetries \cite{greenwood10}. By calculating the position of the peak in Fourier space, we can obtain the real-space lattice constant $\alpha$ of the solid phase in terms of the constants appearing in equation \ref{eq:correlationFourier}, $\alpha = \sqrt[]{{2 \hat{C}_4}/{\hat{C}_2}}$, where we dropped a factor of $4\pi/\,\sqrt[]{3}$ for convenience. Taking the inverse Fourier transform of equation \ref{eq:correlationFourier} returns the correlation to real space giving
	\begin{equation}\label{eq:correlationReal}
	C(|\vec{r}_1-\vec{r}_2|) \approx (-\hat{C}_0-\hat{C}_2 \nabla^2 - \hat{C}_4 \nabla^4)\delta(|\vec{r}_1-\vec{r}_2|)
	\end{equation}
	where $\delta$ is the Dirac delta function.
	
	Next, we define the dimensionless density field $n(\vec{r})=(\rho(\vec{r})-\rho_o)/\rho_o$ which will act as the order parameter of the final derived free energy functional. We also rescale the spatial variable by the lattice constant, $\vec{x} = \vec{r}/\alpha$. Substituting $n$ into equation \ref{eq:ramakF} (truncated to two-point correlation), expanding the nonlinear term in the first integral to fourth order in $n$, and applying one integration on the correlation function obtained in equation \ref{eq:correlationReal} gives
	\begin{multline}\label{eq:PFC_energyFunctional}
	F=\frac{\mathcal{F}}{k_B T \rho_o \alpha^2}= \int d\vec{x} \left\{ \frac{n^2}{2}B^l +\frac{n}{2}B^x(2\nabla^2+\nabla^4)n \right. \\ \left. - \frac{n^3}{6} +  \frac{n^4}{12}\right\}
	\end{multline}
	where we have defined $B^l=1+\rho_o \hat{C}_0$ and $B^x=\rho_o \hat{C}_2^2/4\hat{C}_4$. Equation \ref{eq:PFC_energyFunctional} is the dimensionless free energy functional of the PFC model used in the remainder of this work. The terms of linear or lower order in $n$ in equation \ref{eq:PFC_energyFunctional} were dropped as they do not contribute to the time-evolution PDE given in the next subsection.
	
	The parameter $\Delta B = B^l - B^x$ acts as the effective temperature of the derived PFC model. Returning to the definitions of $B^l$ and $B^x$ and to equation \ref{eq:correlationFourier}, we find that $\Delta B = 1 + \rho_o (\hat{C}_0-\hat{C}_2^2/\hat{C}_4)=1-\rho_o \hat{C}_m$ where $\hat{C}_m$ is the global maximum of the Fourier transformed two-point correlation function. If we fix $\rho_o$ while decreasing $\Delta B$, the peak of the correlation function increases, and vice versa. A higher peak in the correlation function $\hat{C}(k)$ indicates increased preference for the "PFC atoms" to arrange themselves according to the solid phase's reciprocal lattice structure. Thus, decreasing $\Delta B$ is expected to trigger phase transition from liquid to solid. Defining $n_o$ to be the average dimensionless density of the system, one can construct a phase diagram for the presented PFC model in terms of the average density parameter $n_o$ and the effective temperature parameter $\Delta B$ (see Ref.\! \cite{elder04,provatas_PFC} for procedure).
	
	\subsection{Dimensionless time-evolution PDE\label{ssec:time_pde}}
	
	As the PFC order parameter represents an atomic density, the total field $n$ must be conserved as the system is evolved. The time-evolution PDE of the model is thus the Cahn-Hilliard equation (aka. Model B) \cite{hohenberg77}. We start by writing the PDE for the time-evolution of the dimensional density $\rho(\vec{r})$ in terms of the dimensional free energy functional $\mathcal{F}$ of equation \ref{eq:ramakF},
	\begin{equation}\label{eq:pfc_dynamics_dimensionful}
	\frac{\partial \rho}{\partial t} = M \nabla^2 \left(\frac{\delta \mathcal{F}}{\delta \rho}\right) + \nabla \cdot \vec\zeta
	\end{equation}
	where $M$ is a solute mobility parameter and $\nabla \cdot \vec\zeta$ is a noise term representing thermal fluctuations that conserve the total field. $\vec\zeta=(\zeta_x(\vec{r},t),\zeta_y(\vec{r},t))$ is a two-component random vector field, uncorrelated with itself in space and time, and satisfying the fluctuation-dissipation relation \cite{chaikin_condensed,sancho_noise}, expressed as
	\begin{equation}\label{eq:pfc_noise_dimensionful}
	\langle \zeta_i(\vec{r},t),\zeta_j(\vec{r}\,',t') \rangle = -2 k_B T M \delta(\vec{r}-\vec{r}\,')\delta(t-t')\delta_{ij}
	\end{equation}
	where $T$ is the temperature, $k_B$ is the Boltzmann constant, $\delta(\cdot)$ is the Dirac delta function, and $\delta_{ij}$ is the Kronecker delta function. Equation \ref{eq:pfc_noise_dimensionful} is to be interpreted as specifying that each $\zeta_i$ is a random variable uncorrelated with itself and follows a Gaussian distribution with standard deviation $\sigma = \sqrt[]{2k_B T M}$.
	
	To obtain the time-evolution PDE corresponding to our dimensionless free energy functional $F=\mathcal{F}/k_B T \rho_o \alpha^2$ of equation \ref{eq:PFC_energyFunctional}, we again set $n=(\rho-\rho_o)/\rho_o$ and $\vec{x}=\vec{r}/\alpha$, giving
	\begin{equation}\label{eq:pfc_dynamics_dimensionless}
	\frac{\partial n}{\partial t} = \Gamma \nabla^2 \left(\frac{\delta F}{\delta n}\right) + \nabla \cdot \vec\xi
	\end{equation}
	where $\Gamma = k_B T M/\rho_o $ is the dimensionless solute mobility parameter, and the dimensionless noise term satisfies
	\begin{equation}\label{eq:pfc_noise_dimensionless}
	\langle \xi_i(\vec{x},t),\xi_j(\vec{x}\,',t') \rangle = - N_a^2 \delta(\vec{x}-\vec{x}\,')\delta(t-t')\delta_{ij}
	\end{equation}
	where $N_a^2=2\Gamma/\rho_o \alpha^2$ \cite{kocher16}. Evaluating the functional derivative in equation \ref{eq:pfc_dynamics_dimensionless} gives the time-evolution PDE for the dimensionless density $n(\vec{x})$, written as
	\begin{equation}\label{eq:pfc_dynamics_dimensionless_final}
	\frac{\partial n}{\partial t} = \Gamma \nabla^2 \left[(B^l +B^x (2\nabla^2+\nabla^4))n -\frac{n^2}{2}+\frac{n^3}{3}\right] + \nabla \cdot \vec\xi
	\end{equation}

	It is instructive to discuss the dimensionless  standard deviation of the noise, $N_a$. For a known  $\Gamma$, one could attempt to match it to a specific real material's values at the reference liquid density $\rho_o$ and the model's dimensional lattice constant $\alpha$. However, it has been shown by Kocher et. al \cite{kocher16} that  $\rho_o\alpha^2$ (and hence $N_a$) must be chosen as a function of the PFC model's $B^l$ and $B^x$ parameters (also dimensionless) to ensure proper behavior of capillary fluctuations of a solid-liquid interface. This essentially reflects the crudeness of the PFC model's approximations, and ignorance of the precise coarse graining volume of our coarse grained PFC theory. Ref.~\cite{kocher16} also shows that a cutoff must be applied to the noise  spectrum: noise modes with wavenumber $k>2\pi/a$ in Fourier space must be set to zero, where $a$ is the dimensionless lattice constant (obtained by minimizing the dimensionless free energy functional for a solid bulk, not to be confused with the dimensional lattice constant $\alpha$). This cutoff can be understood as eliminating unphysical fluctuations on scales smaller than the lattice separation, as these fluctuations would have already been accounted for in obtaining the CDFT used in subsection \ref{ssec:f_func} to derive the PFC  free energy functional. It can also be understood from a numerical perspective \cite{plapp11}: not implementing such a cutoff causes the atomic-scale dynamics of the simulated model to strongly depend on the discretization scheme used, due to more noise modes being available for a finer grid discretization.
	
	Though we have scaled out the explicit temperature dependence form our model and from equation \ref{eq:pfc_noise_dimensionless}, we require that the equilibrium probability distribution of states of our system continue obeying the Boltzmann distribution \cite{sancho_noise}. This is ensured through the fluctuation-dissipation theorem in equation \ref{eq:pfc_noise_dimensionless}, which in its dimensionless form can now be considered as having a dimensionless fluctuation temperature $T_{r}$, defined through $N_a^2\approx 2\Gamma T_{r}$. The relation is only approximate due to the cutoff applied to the noise's Fourier modes. In this work, we assume that $T_{r}$ is used in calculating quantities related to the fluctuation-driven dynamics of the system, such as the Boltzmann factor $\exp(-E/T_r)$ that gives the probability of a state of dimensionless energy $E$ relative to the probability of a state of zero energy. This fluctuation temperature should not be confused with either the dimensional $T$ that was scaled out of the PFC free energy, or $\Delta B$ which is normally considered the model's effective temperature due to its role in determining the equilibrium phase diagram for the model.  $T_{r}$ and $\Delta B$ are effectively coupled by following the values of $N_a$ versus $\Delta B$ found in Ref.~\cite{kocher16}, and care should be taken when using these separately in equations that require a temperature value or dependence. 
	
	%%%%%%%%%%%%%%%%%%%%%%%%%%%%%%%%%%%%%%%%%%%%%%%%%%%%%%%%%%%%%%%%%%%%%%%%%%%%%%%%%%%%
	
	\section{Classical nucleation theory\label{sec:cnt}}
	
	\subsection{Work of formation\label{ssec:work_form}}
	
	Consider a pure liquid material capable of undergoing phase transition to a crystalline solid state. The atoms of the disordered liquid phase undergo constant thermal fluctuations, occasionally stochastically arranging into a structure resembling a small grain of the crystalline solid. These grains can then proceed to either dissolve back into disordered liquid due to further fluctuations, or continue growing as more atoms attach to the original structure. If the liquid is above its melting temperature, the grains will always eventually dissociate into component liquid atoms. However, below the melting temperature, whether the grains are stable to fluctuations and continue growing or not depends on their size, density, shape, as well as other factors. Classical nucleation theory (CNT) \cite{hoyt_phasetransf,sear07} attempts to predict the rate of appearance of stable solid grains under the simplest possible assumptions for factors determining their stability.
	
	In CNT, the interior of a solid grain is treated as consisting of bulk solid, with a sharp interface separating it from the surrounding liquid phase. The `work of formation' $W$ is the free energy required to form such a grain from the original liquid phase. $W$ consists of a bulk term, which represents the difference between the free energies of the solid and liquid phase, as well as a surface term representing the energy penalty for the existence of an interface. Assuming a circular two-dimensional grain, we write
	\begin{equation}\label{eq:nuc_work_circle}
	W(R) = \pi R^2\Delta G + 2\pi R \gamma
	\end{equation}
	where $R$ is the radius of the grain, $\Delta G$ is the difference in local free energy density between the initial and final phases, and $\gamma$ is the interfacial energy density. When the system is below its melting point, the solid phase is favored, leading to a negative $\Delta G$. The nucleation energy barrier is given by the maximum of equation \ref{eq:nuc_work_circle}, with value $W^*$ and position $R^*$ found to be
	\begin{equation}\label{eq:nuc_work_circle_max}
	W^* = -\pi\frac{\gamma^2}{\Delta G}, \qquad R^* = -\frac{\gamma}{\Delta G}
	\end{equation}
	and a critical nucleus is then defined to be a grain of radius $R^*$, while larger grains are termed post-critical nuclei.
	
	CNT assumes that atoms in a grain have their mass evenly distributed throughout the grain. We can thus rewrite equation \ref{eq:nuc_work_circle} in terms of the number of atoms in the grain, giving
	\begin{equation}\label{eq:nuc_work_g}
	W(g) = vg\Delta G + sg^{1/2} \gamma
	\end{equation}
	where $g$ is the number of atoms, and $v$ and $s$ are, respectively, the area and interfacial length of an effective two-dimensional grain consisting of 1 atom ($g=1$). The critical nucleus' number of atoms $g^*$ is then found similarly as in equation \ref{eq:nuc_work_circle_max}, giving
	\begin{equation}\label{eq:nuc_crit_num_g}
	g^* = \left(-\frac{s\gamma}{2v\Delta G}\right)^2
	\end{equation}

	Note that in this work we only consider homogeneous nucleation for simplicity, though the case of heterogeneous nucleation can also be examined with the same formalism by modifying the form of $\gamma$.
	
	\subsection{Time-dependent nucleation rate\label{ssec:nuc_timedep}}
	
	A rapidly quenched liquid system does not instantaneously exhibit its maximum possible nucleation rate, instead requiring a finite amount of time, known as the `incubation time', for the nucleation rate to approach the late-time `steady-state' rate. The time-dependent nucleation rate in the nucleating liquid phase is defined to be the flux in size-space $g$ of grains at the critical nucleus size $g^*$. An approximate form for this rate is obtained by Shi, Seinfeld and Okuyama \cite{shi90} by solving a Fokker-Planck equation for the grain size distribution using singular perturbation methods, under the assumption that critical nuclei consist of a large number of atoms ($g^*>>1$), an assumption known to be true for most physical systems. This rate is
	\begin{equation}\label{eq:nuc_crit_flux}
	J^*(t)=J_{ss} \exp\left[-\exp\left(-2\frac{t}{\tau}+2\lambda\right)\right]
	\end{equation}
	where $J_{ss}$ is the steady state nucleation rate, and $\tau$ and $\lambda$ are values that depend on $g^*$ and determine the incubation time. $J_{ss}$ is given as \cite{hoyt_phasetransf,sear07}
	\begin{equation}\label{eq:nuc_rate_ss}
	J_{ss} = Zj^*\rho(g^*)= Zj^* \rho_1\exp\left(-\frac{W^*}{k_B T}\right)
	\end{equation}
	where $j^*$ is the rate of single-atom attachment to an exactly critical nucleus, $\rho_1$ is the number density of  atoms in the liquid, and $Z$ is the Zeldovich factor. $\rho_1$ is taken to be constant, under the assumption of no external mass-exchange. $Z$ and $j^*$ are expected to scale in two dimensions as
	\begin{multline}\label{eq:nuc_Z_scaling}
	Z \propto \left(-\frac{1}{k_B T}\frac{\partial^2 W}{\partial g^2}\Bigr|_{g=g^*}\right)^{1/2} \\ \propto \left(\frac{(g^*)^{-3/2}\gamma}{k_B T}\right)^{1/2}  \propto \frac{(-\Delta G)^{3/2}}{\gamma  (k_B T)^{1/2}}
	\end{multline}
	\begin{equation}\label{eq:nuc_attachmentrate}
	j^* \propto (g^*)^{1/2} k_B T \exp\left(-\frac{\Delta G_A}{k_B T}\right)
	\end{equation}
	where $\Delta G_A$ is the activation energy needed for an atom to cross the liquid-solid interface to attach to the crystal grain. Further, $\tau$ and $\lambda$ are calculated to scale in two dimensions as
	\begin{equation}\label{eq:nuc_tau_scaling}
	\tau \propto \frac{1}{Z^2 j^*}
	\end{equation}
	\begin{equation}\label{eq:nuc_lambda_scaling}
	\lambda \propto  (g^*)^{-1/2} -1 + \ln\left(Zg^*(1-(g^*)^{-1/2})\right) +\ln\left(2\sqrt{\pi}\right)%O(\ln(g^*))
	\end{equation}

	It proves to be numerically (and experimentally) easier to calculate the number of post-critical nuclei in a system than it is to directly calculate their rate of appearance. Hence, we derive the time-dependent number density of post-critical nuclei by taking the integral of equation \ref{eq:nuc_crit_flux}. This gives
	\begin{equation}\label{eq:nuc_crit_numdens}
	I^*(t)=\int_0^t J^*(s)ds=- \frac{J_{ss} \tau}{2} \mathrm{Ei}\left[-\exp\left(-2\frac{t}{\tau}+2\lambda\right)\right]
	\end{equation}
	where $\mathrm{Ei}(.)$ is the exponential integral function, defined as
	\begin{equation}
	\mathrm{Ei}(x)=\int_{-\infty}^x\frac{e^s}{s}ds
	\end{equation}

\begin{figure}[h]
		\subfloat[\label{fig:nuc_rates_fixedinc_a}]{%
			\includegraphics[width=0.45\textwidth]{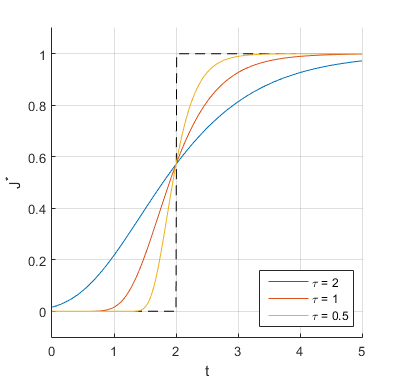}
		}
	
		\subfloat[\label{fig:nuc_rates_fixedinc_b}]{%
			\includegraphics[width=0.45\textwidth]{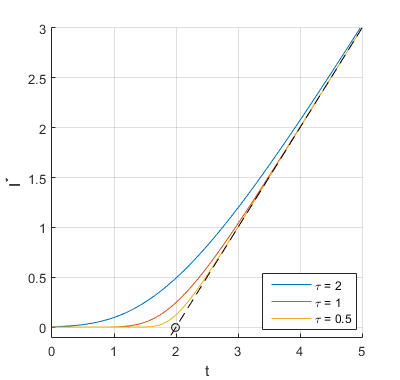}
		}
		\caption{(a) Sketch of $J^*(t)$ for $J_{ss}=1$ and $t^*=2$ fixed while $\tau$ varies. The black curve shows the step function for $\tau \rightarrow 0$. (b) Sketch of $I^*(t)$ for $J_{ss}=1$ and $t^*=2$ fixed while $\tau$ varies. The black curve shows the asymptote line, and its intersection with the horizontal axis is shown by the black circle.}\label{fig:nuc_rates_fixedinc}
	\end{figure}

	It is not immediately clear from the forms of equations \ref{eq:nuc_crit_flux} and \ref{eq:nuc_crit_numdens} what value should be considered the incubation time, as both $\tau$ and $\lambda$ affect the time needed to reach steady state nucleation rate $J_{ss}$. In this work, we will define the incubation time geometrically, in a manner similar to experimental works such as \cite{legoues84}. As $t \rightarrow \infty$, we calculate that $I^*(t)$ asymptotes to a line given by
	\begin{equation}\label{eq:nuc_crit_numdens_asymptote}
	I^*(t) \approx J_{ss} ( t - \tau(\lambda+\gamma_{e}/2) )
	\end{equation}
	where $\gamma_{e}\approx 0.5772$ is the Euler-Mascheroni constant, a well-known mathematical constant that we obtain when solving for the asymptotic behavior of equation \ref{eq:nuc_crit_numdens}. The intercept of this line with the horizontal axis is then taken to be the incubation time, written as
	\begin{equation}\label{eq:nuc_incubationtime}
	t^*=\tau(\lambda+\gamma_{e}/2)
	\end{equation}
	Figure \ref{fig:nuc_rates_fixedinc} sketches $J^*(t)$ and $I^*(t)$ for a fixed incubation time $t^*$ as $\tau$ and $\lambda$ are varied according to equation~\ref{eq:nuc_incubationtime}.
	
	Finally, we estimate the predicted scalings of $J_{ss}$ and $t^*$ by combining equations \ref{eq:nuc_work_circle_max}, \ref{eq:nuc_crit_num_g}, \ref{eq:nuc_rate_ss}, \ref{eq:nuc_Z_scaling}, \ref{eq:nuc_attachmentrate}, \ref{eq:nuc_tau_scaling}, \ref{eq:nuc_lambda_scaling}, and \ref{eq:nuc_incubationtime}. We find
	\begin{equation}\label{eq:nuc_Jss_scaling}
	J_{ss} \propto (-\Delta G \, k_B T)^{1/2}\exp\left(-\frac{\Delta G_A}{k_B T}\right)\exp\left(-\frac{\pi}{k_B T}\frac{\gamma^2}{(-\Delta G)}\right)
	\end{equation}
	\begin{equation}\label{eq:nuc_tinc_scaling}
	t^* \propto \left[\frac{1}{(-\Delta G)}+K\frac{\gamma}{(\Delta G)^2}\right]\exp\left(+\frac{\Delta G_A}{k_B T}\right)
	\end{equation}
	where we have expressed the scalings in terms of the local free energy density difference $\Delta G$ between the solid and liquid phase, the interfacial energy density $\gamma$ of a grain, the temperature factor $k_B T$, and the atomic-attachment activation energy $\Delta G_A$. We also defined
	\begin{multline}
	K = \lambda - (g^*)^{-1/2} +\gamma_e/2 \\ \propto \gamma_e/2-1+ \ln\left(2\sqrt{\pi}\right)+\ln\left(\frac{\gamma}{(-\Delta G \, k_B T)^{1/2}}  \left(1-\frac{(-\Delta G)}{k_B T}\right)  \right), 
	%K = \lambda - (g^*)^{-1/2}
	\end{multline}
	which we will take to be approximately constant due to the slow variation of the logarithmic function. The sign of $K$ is seen to be positive if $\lambda$ is positive, under the assumption that $g^*>>1$. In turn, we see from the time-dependent rate equation \ref{eq:nuc_crit_flux} that $\lambda$ must be positive for the rate to be approximately zero at $t=0$. As such, we will assume $K>0$. These assumptions will be useful in the discussion of our results in section \ref{sec:res}.
	
	\subsection{Modeling nucleation in the PFC model\label{ssec:nuc_pfc}}
	
	To compare nucleation in the PFC model to the predictions of CNT, we will obtain the number density of post-critical nuclei $I^*(t)$ computed in different simulated PFC systems. This number density will be used to calculate $J_{ss}$ and $t^*$ geometrically as described in the previous section. We will then examine the scaling of these two values in the PFC model and compare them to equations \ref{eq:nuc_Jss_scaling} and \ref{eq:nuc_tinc_scaling}. To do so, we will assume that the physical values appearing in these two equations ($\gamma$, $\Delta G$, $\Delta G_A$, and $k_B T$) have equivalent or effective dimensionless counterparts in our dimensionless PFC model. Section \ref{sec:res} will briefly describe the numerical methods used to collect our PFC data, and showcase the corresponding results and comparison to CNT. In the remainder of this section, we preemptively describe the difficulties we expect to encounter in this comparison due to the assumptions made by CNT in its derivations.
	
	The first difficulty will relate to the finite size of the simulated systems. As CNT assumes grains do not interact and also assumes the number density of single atoms (equivalently, homogeneous nucleation sites) is constant through time, its predictions are expected to only hold in early times for the PFC model simulations, before a significant fraction of the liquid phase has transitioned to the solid phase. For this reason, we will be unable to definitively ascertain that an $I^*(t)$ obtained from  simulations has reached its predicted late-time asymptotic value before it tapers off due to finite size limits. As such, calculating $J_{ss}$ and $t^*$ geometrically from the presumed asymptote line are only guaranteed to provide lower bounds for these values, rather than exact results.
	
	Another difficulty is due to the approximation used to obtain equation \ref{eq:nuc_crit_flux} for the time-dependent nucleation rate. A close examination of the perturbative approach used by the authors of Ref.~\cite{shi90} to derive the time-dependent rate reveals an assumption that the number of atoms in a critical nucleus is large, $g^*>>1$. For numerical efficiency reasons, our simulations of the PFC model will be using parameters that lead to critical nuclei consisting of a small number of density peaks, corresponding to few atoms: $g^*\approx 5$. As such, the results of the singular perturbation derivation will likely not hold exactly, though we expect that the qualitative features of the predicted time dependence will still be present.
	
	A third and more subtle difficulty is found in CNT's use of a single variable to describe grains, the number of atoms $g$ in a grain. In both the PFC model and other nucleating models, this can prove to be an oversimplification, as shape, density, and interface width of the grains can vary independently during the formation process. See for example Ref.~\cite{lutsko15} where nucleation in globular protein systems is assumed to depend both on interior density and radius of the grains. It is then unclear whether the definitions of $\gamma$ and $\Delta G$ are sound for small pre-critical grains, as $\gamma$ assumes a sharp interface (or at least an interface width much smaller than a bulk solid grain's width) and $\Delta G$ assumes inner grain density equal to the final solid bulk density. In addition, due to vibrational-timescale fluctuations being averaged out in the PFC model, it is unclear whether the assumption of single-atom attachment rate in the form of equation \ref{eq:nuc_attachmentrate} is a reasonable approximation, and no direct equivalent to the activation energy $\Delta G_A$ is available in this model. Furthermore, as the PFC model's systems consist of a continuous density field rather than discrete atoms, it is feasible that the formation of grains involves fluctuations in the field that do not follow the expected lattice structure at early times, before the grains stabilize. See for example Ref.~\cite{toth11} where the authors observe what appears to be amorphous structure appearing in the PFC model preceding a crystalline phase, though they are unable to conclude whether this structure represents a separate amorphous phase or very small and tightly packed crystal grains. As part of the discussion of our results, we will attempt to numerically calculate an approximation for the form of the critical nucleus in the PFC model. We will also examine the behavior of the phase-field (smoothed density field) during the early formation stage of the grains. These undertakings will be used as guides to assess whether CNT assumptions are reasonable for the PFC model. We will thus continue assuming the definitions of $\gamma$ and $\Delta G$ hold, at least in some approximate manner, when examining the scalings of $J_{ss}$ and $t^*$.
	
	The final hurdle relates to the two different temperature parameters that are defined in the PFC model: the effective temperature $\Delta B$ obtained from the parameters in the model's free energy functional in equation \ref{eq:PFC_energyFunctional}, and the fluctuation temperature $T_r$ that follows from the fluctuation-dissipation theorem inherent in defining fluctuations in equation \ref{eq:pfc_noise_dimensionless}.	While authors of Ref.~\cite{kocher16} show that these two temperatures need to be coupled to correctly reproduce capillary fluctuations, it isn’t clear that this holds in the context of nucleation. Furthermore, there is no definitive way to decide which temperature dependencies in quantities in the CNT scaling predictions of equations \ref{eq:nuc_Jss_scaling} and \ref{eq:nuc_tinc_scaling} correspond to each of $\Delta B$ and $T_r$. We therefore choose to consider these temperatures separately.	In this work, we make the following working assumptions related to temperature dependence: Factors of $k_B T$ appearing in the exponential terms $\exp(-\Delta G_A/k_B T)$ and $\exp(-W^*/k_B T)$ are taken to correspond to $T_r$ in the dimensionless PFC model, as these exponential terms are based on the Boltzmann distribution arguments discussed in subsection~\ref{ssec:time_pde} (recall that the `true' dimensional $k_B T$ has been scaled out in our PFC model, leading all energies used in this model to be dimensionless). Additionally, we assume  that the temperature dependence of $\Delta G$ is reflected only in its dependence on $\Delta B$, and can be approximated from the difference of the local free energy densities of solid and liquid bulks obtained using the standard one-mode approximation for the PFC model \cite{provatas_PFC,provatas07}. Further, interfacial energies of stable interfaces in the PFC model are known \cite{toth10} to decrease slowly as $\Delta B$ increases, and thus we take $\gamma$ to vary as such. Finally, for physical systems such as water below its freezing point \cite{jeffery97}, $\Delta G_A$ is estimated to decrease with increasing temperature. We thus assume that, if its effects are present in the PFC model, $\Delta G_A$ decreases with one or both of the temperature parameters, though its exact dependence is unknown.

	 We will attempt to evaluate these above assumptions based on the results we obtain.
	
	%%%%%%%%%%%%%%%%%%%%%%%%%%%%%%%%%%%%%%%%%%%%%%%%%%%%%%%%%%%%%%%%%%%%%%%%%%%%%%%%%%%%
	
	\section{Results and discussion\label{sec:res}}
	
	\subsection{Nucleation rates and incubation times Vs. $\Delta B$\label{ssec:DB_varies}}
	
\begin{figure*}[!ht]
		\centering
		\begin{tabular}{cccc}
			\subfloat[]
			{
				\includegraphics[width=0.45\textwidth]{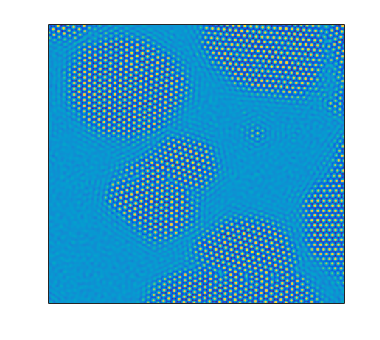}
			}
			&\subfloat[]
			{
				\includegraphics[width=0.45\textwidth]{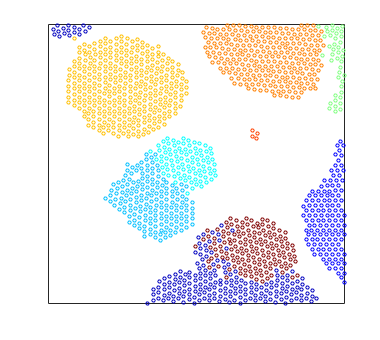}
			}
		\end{tabular}
		\caption{(a) shows a snapshot of the PFC density field $n$ in a small area of a simulation domain. (b) shows the corresponding detected peaks, grouped into clusters shown by color. Note that the method used to distinguish clusters does not always perfectly separate impinging clusters at their lattice boundaries. This is not an issue as we are interested in the number of such distinguishable clusters, rather than their exact final size.}\label{fig:num_clustexample}
	\end{figure*}

	The PFC model simulations used were implemented with grid size $1024\times1024$, dimensionless time step $dt=1$, and dimensionless space step $dx=a/8\approx0.91$ where $a$ is the lattice spacing. Periodic boundary conditions where applied. The time-stepping of the PFC model's PDE was done using a semi-implicit Fourier space method. The simulation code was implemented in MATLAB and ran on a single GPU device. To calculate the number density of post-critical nuclei $I^*(t)$, the local peaks of the simulated periodic PFC density field were grouped into clusters, and clusters observed to continue growing after initial detection where counted as post-critical nuclei. Figure \ref{fig:num_clustexample} shows a zoomed-in snapshot of a simulation run's PFC density field and corresponding detected peaks grouped into clusters.
	
	A total of six data sets where obtained for this subsection. Each data set consists of the averaged results of 50 to 150 simulation runs. All data sets had fixed PFC model parameters $n_o=0.207$, $B^x=0.4$, and $N_a=0.040$. The effective PFC temperature $\Delta B$ increases between the data sets, chosen to be $\Delta B=0.16500 + 0.00025\epsilon$ for $\epsilon$ an integer from 0 to 5 corresponding to the six sets in order from first to last. The noise amplitude $N_a$ was chosen approximately within the recommended range corresponding to the other parameters as given in Ref.~\cite{kocher16}, bearing in mind that these authors' results show that the variation range of $\Delta B$ between our data sets is too small to warrant a change in $N_a$ between the sets. Table \ref{tab:params_epsilon} lists the data sets' parameters and number of runs.
	
	\begin{table}[]
		\centering
		\begin{tabular}{c|cccc|c}
			$\epsilon$&  $\Delta B$&  $B^x$&  $n_o$&  $N_a$&Number of runs\\ \hline
			0&  0.16500&  0.4&  0.207&  0.040&100\\
			1&  0.16525&  0.4&  0.207&  0.040&50\\
			2&  0.16550&  0.4&  0.207&  0.040&50\\
			3&  0.16575&  0.4&  0.207&  0.040&50\\
			4&  0.16600&  0.4&  0.207&  0.040&150\\
			5&  0.16625&  0.4&  0.207&  0.040&100
		\end{tabular}
		\caption{PFC model parameters used to generate data sets $\epsilon=0$ to $\epsilon=5$. Also given is the number of simulation runs used to obtain the averaged results for each data set.}
		\label{tab:params_epsilon}
	\end{table}
	
	The parameters chosen for the aforementioned  runs place the simulated systems below the solidus (at $\Delta B \approx 0.1685$ for the chosen $n_o$ and $B^x$) on the phase diagram, and above the instability curve (at $\Delta B \approx 0.1642$). The range of $\Delta B$ was chosen such as to allow appreciable amounts of nucleation to take place in a reasonable amount of computational time, while remaining above the instability curve. We note that this range is small relative to the range between the instability curve and the solidus, and that the range lies closer to the instability curve than to the solidus, indicating that the systems are greatly undercooled below their freezing point for nucleation to occur at noticeable rates. While this might be a result of making the PFC model's equations dimensionless, there are some relevant physical systems exhibiting such homogeneous nucleation behaviour. For example, in the absence of heterogeneous nucleation, water is known to remain liquid at temperatures of $235K$ and below \cite{mason58,jeffery97}. See also \cite{hoyt_phasetransf}, where an estimate for the variation of $J_{ss}$ with temperature is obtained using realistic values for an alloy. This estimate predicts that the homogeneous nucleation rate is undetectable before a specific temperature more than $100K$ below the alloy's freezing point, yet the rate rapidly increases beyond that specific undercooling temperature, similar to the behavior our results will show.
	
\begin{figure}[!h]
		\centering
		\includegraphics[width=0.45\textwidth]{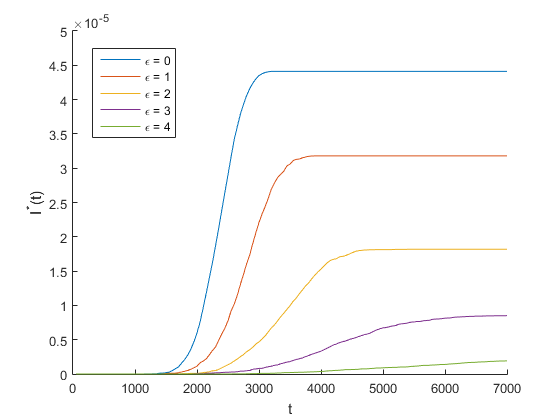}
		\caption{The post-critical nuclei densities for data sets $\epsilon=0$ to $\epsilon=4$. Time is in units of $d t$.}\label{fig:res_I_datasets}
	\end{figure}

	Figure \ref{fig:res_I_datasets} plots the post-critical nuclei densities $I^*(t)$, corresponding to the quantity in equation \ref{eq:nuc_crit_numdens}, for the first 5 sets. The sixth set, with $\epsilon=5$, is not visible on that figure's scale. We observe that the early portion of these curves resembles the form predicted by CNT, such as in figure \ref{fig:nuc_rates_fixedinc_b}, except that it is unclear whether the linearly increasing parts of these curves reach the true asymptote before tapering off to a constant value at late times due to the system fully transitioning to solid.
	
	\begin{figure}[!h]
		\subfloat[]{%
			\includegraphics[width=0.4\textwidth]{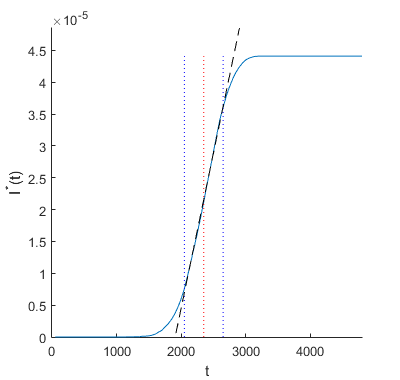}
		}
	
		\subfloat[]{%
			\includegraphics[width=0.4\textwidth]{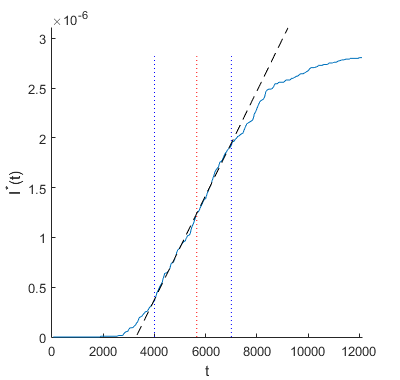}
		}
		\caption{The geometric construction used to obtain $J_{ss}$ and $t^*$. Time is in units of $d t$. The vertical blue dotted lines indicate the time range encompassing the asymptote. The dashed black line is the linear fit over that time range. Also shown as a vertical red dotted line is the time at which 10\% of the initial liquid volume has solidified. (a) Data set $\epsilon=0$, giving $J_{ss}=(4.9\pm 0.1)\times10^{-8}$ and $t^*=1910\pm 70$. (b) Data set $\epsilon=4$, giving $J_{ss}=(5.23\pm 0.08)\times10^{-10}$ and $t^*=3280\pm 90$. The errors on these values are from the uncertainty on the linear fit.}\label{fig:res_I_datasets_examples}
	\end{figure}
	
	We obtain the steady-state rate of nucleation $J_{ss}$ and the incubation time $t^*$ by the geometric construction described in subsection~\ref{ssec:nuc_timedep}, taking as an assumption that the linear part of each data set's $I^*(t)$ corresponds to the CNT-predicted asymptote. As a check for whether the assumption of asympoticity is sound, we also obtain for each data set the fraction of the initial liquid volume that transitioned to solid as a function of time. We find that in all the data sets, only 10\% of the total liquid volume has solidified by the time approximately half of all post-critical nuclei have appeared. As the $I^*(t)$ curves have already entered the linear regime before that time, this suggests that the liquid has not yet been significantly depleted in the time range we assume to correspond to the asymptote. Figure \ref{fig:res_I_datasets_examples} demonstrates the geometric constructions used to obtain $J_{ss}$ and $t^*$ for two of the data sets.%, as well as the time at which 10\% of the liquid volume has solidified.
	
	\begin{figure}[!h]
		\subfloat[]{%
			\includegraphics[width=0.45\textwidth]{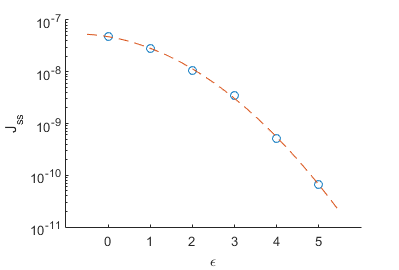}\label{fig:res_Jss}
		}
		
		\subfloat[]{%
			\includegraphics[width=0.45\textwidth]{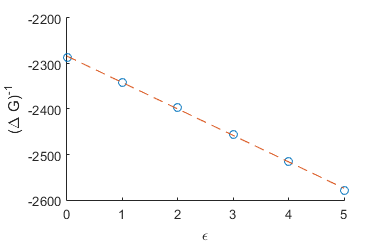}\label{fig:DG_tempdep}
		}
		\caption{(a) Plot of $J_{ss}$ as a function of  effective temperature $\Delta B$ for data sets $\epsilon=0$ to $\epsilon=5$ (blue circles), along with quadratic fit in semi-log space (red dashed line) showcasing faster than exponential decrease. Error bars are not visible at this scale. (b) Plot of $(\Delta G)^{-1}$ versus $\Delta B$, over the range $\Delta B$ for the data sets $\epsilon=0$ to $\epsilon=5$. Blue circles are calculated values, red dashed line is a linear fit.}
	\end{figure}
	
	Figure \ref{fig:res_Jss} plots $J_{ss}$ versus $\Delta B $ for the six data sets, on a semi-log plot. Also shown is a quadratic fit in semi-log space to demonstrate faster than exponential decrease of $J_{ss}$ as effective temperature $\Delta B$ increases. To compare this change in rate to equation \ref{eq:nuc_Jss_scaling}, we only consider the effect of the exponential terms in that equation. As discussed in subsection~\ref{ssec:nuc_pfc}, we replace $k_B T$ by the dimensionless fluctuation temperature $T_r=N_a^2/2$, held constant since $N_a$ does not vary between the data sets. We also take $\Delta G_A$ and $\gamma$ (which are dimensionless due to the physical $k_BT$ having been scaled out) to be positive and decreasing as the effective temperature $\Delta B$ increases. We then turn to $\Delta G$ to try and explain the observed change in $J_{ss}$. Figure \ref{fig:DG_tempdep} plots $(\Delta G)^{-1}$ over the range of $\Delta B$ corresponding to the data sets, with $\Delta G$ estimated to be the difference between the solid and liquid bulk free energy densities of the PFC model, obtained from the standard one-mode approximation for the model. We observe that this plot varies approximately linearly with $\Delta B$, meaning $\Delta G$ can only account for at most exponential decrease, not faster than exponential. This suggests either that the CNT definitions of $\Delta G$, $\gamma$ or $\Delta G_A$ are inadequate for the PFC model, as mentioned in subsection~\ref{ssec:nuc_pfc}, or that the linear fit in the geometric construction used to obtain $J_{ss}$ was not at the true asymptote, which might have resulted in a lower bound for $J_{ss}$ that is less accurate for the data sets of lower $\epsilon$.
	
	We believe the discrepancy mentioned in the previous paragraph is due to CNT's assumption that the surface energy density $\gamma$ of a small forming grain is equivalent to the that of a much larger solid bulk's interface. This assumption was what led us to take $\gamma$ as slowly decreasing with increasing model temperature, since that is the behavior of a solid bulk's interfacial energy density both in the PFC model and in real crystalline materials. However, this assumption contradicts the existence of the instability curve of the model: in particular, equation \ref{eq:nuc_work_circle_max} for the nucleation energy barrier $W^*$ implies that $\gamma$ must vanish for the nucleation barrier to vanish (since $\Delta G$ remains finite at the instability limit). Thus, the existence of the unstable region on the PFC model's phase diagram (where by necessity the nucleation barrier must vanish) implies $\gamma$ for small grains must {\it increase} from zero at the instability curve to a higher value as effective temperature $\Delta B$ increases. The value of $\gamma$ for a nucleating grain then can not be the same as the interfacial energy density in a grain with a larger bulk, likely due to $\gamma$ including energetic contributions from effects such as high interface curvature which are not as significant for interfaces in larger grains. Taking as a new assumption that $\gamma$ of a small forming grain (near the instability curve) increases with temperature leads to qualitative agreement between equation \ref{eq:nuc_Jss_scaling} and the faster that exponential decrease of $J_{ss}$ seen in figure \ref{fig:res_Jss}.
	
	We note that experiments for physical materials qualitatively support the predicted form of the $J_{ss}$ plot we obtained. For example, see figure 6 in Ref.~\cite{jeffery97} and figure 8a in Ref.~\cite{legoues84}, which respectively show steady state homogeneous nucleation rates for water and a CuCo alloy. These experiments observe a faster than exponential decrease as temperature increases on the high-temperature parts of these plots. A quantitative fit to experimental results using the PFC model is still an active area of research and is beyond the scope of this work. Further, we are unable to access the low-temperature regime of experimental plots using the PFC model described in this work, as the decrease in nucleation rate as temperature decreases requires a model with a temperature-dependent mobility.
	
\begin{figure}[h]
		\centering
		\includegraphics[width=0.45\textwidth]{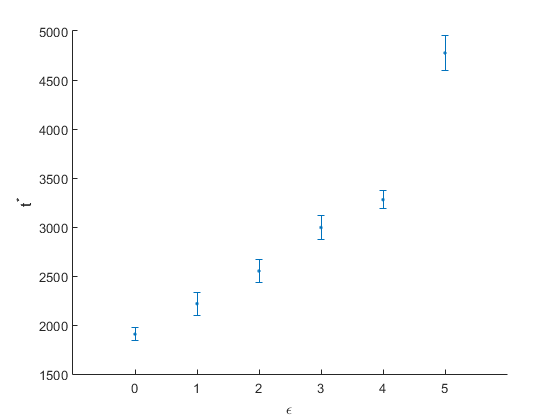}
		\caption{Plot of incubation time $t^*$ for the PFC model data sets $\epsilon=0$ to $\epsilon=5$. Error bars were obtained from the uncertainty on the linear fit to the asymptote, as mentioned in figure \ref{fig:res_I_datasets_examples}.}\label{fig:res_tinc}
	\end{figure}

	Figures \ref{fig:res_tinc} plots $t^*$ for the aforementioned PFC data sets. We observe that $t^*$ increases with effective temperature $\Delta B$. Comparing to equation \ref{eq:nuc_tinc_scaling}, the exponential term in that equation can not be the source of that increase under the assumptions that $k_B T$ is replaced by $T_r=N_a^2/2$ and $\Delta G_A$ decreases with the effective temperature parameter of the dimensionless PFC model. As figures \ref{fig:res_I_datasets} and \ref{fig:res_I_datasets_examples} demonstrate, for our datasets, $J^*(0)\approx 0$ and thus we can safely assume that $K>0$ in equation \ref{eq:nuc_tinc_scaling}. Using our new assumption that $\gamma$ increases with temperature as well as the variation of $(\Delta G)^{-1}$ from figure \ref{fig:DG_tempdep}, we can then conclude that the pre-exponential terms in equation \ref{eq:nuc_tinc_scaling} agree with the observed increase of $t^*$ with $\Delta B$, though no exact fit is attempted. This increase also agrees with the predictions for physical materials (for example, see the TTT curves of figure 1 in Ref.~\cite{legoues84}), again for temperatures where temperature-dependent mobility is negligible.
	
	\subsection{Nucleation rates and incubation times Vs. $N_a$\label{ssec:Na_varies}}
	
	The data sets of subsection \ref{ssec:DB_varies} allowed the comparison of nucleation in the PFC model to CNT as the effective temperature $\Delta B$ varied. The noise amplitude $N_a$ was taken to be coupled to the effective temperature $\Delta B$ as prescribed in Ref.~\cite{kocher16}, which led to $N_a$ being approximately constant over the small chosen range of $\Delta B$. It is of interest to examine the behavior of nucleation as a function of noise amplitude alone, to see whether our assumptions about the temperature dependence of exponential factors in equations \ref{eq:nuc_Jss_scaling} and \ref{eq:nuc_tinc_scaling} were warranted. In this section, we relax the requirement found in Ref.~\cite{kocher16} on the noise amplitude parameter $N_a$, allowing it to be varied independently of $\Delta B$ over multiple new data sets. We stress that in these new data sets, $J_{ss}$ and $t^*$ curves are not expected to vary as would be predicted for a physical system, as decoupling the choice of $N_a$ from $\Delta B$ effectively leads to the thermal fluctuations being unphysical at the scale of the capillary length, as found in Ref.~\cite{kocher16}.
	
	We generate six new data sets with fixed model parameters $n_o=0.207$, $B^x=0.4$, and $\Delta B=0.1650$, and with increasing noise amplitude $N_a=0.030+0.002\kappa$ for $\kappa$ an integer from 0 to 5 corresponding to the six sets respectively. Table \ref{tab:params_kappa} lists the parameters of the new data sets and number of runs for each.
	
	\begin{table}[]
		\centering
		\begin{tabular}{c|cccc|c}
			$\kappa$&  $\Delta B$&  $B^x$&  $n_o$&  $N_a$&Number of runs\\ \hline
			0&  0.16500&  0.4&  0.207&  0.030&100\\
			1&  0.16500&  0.4&  0.207&  0.032&150\\
			2&  0.16500&  0.4&  0.207&  0.034&50\\
			3&  0.16500&  0.4&  0.207&  0.036&50\\
			4&  0.16500&  0.4&  0.207&  0.038&50\\
			5&  0.16500&  0.4&  0.207&  0.040&100
		\end{tabular}
		\caption{PFC model parameters used to generate data sets $\kappa=0$ to $\kappa=5$. Also given is the number of simulation runs used to obtain the averaged results for each data set.}
		\label{tab:params_kappa}
	\end{table}
	
\begin{figure}[h]
		\centering
		\includegraphics[width=0.45\textwidth]{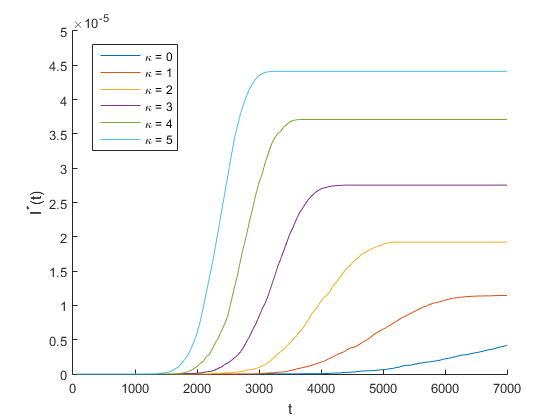}
		\caption{Post-critical nuclei densities for data sets $\kappa=0$ to $\kappa=5$. Time is in units of $d t$.}\label{fig:res_I_datasets_newnoise}
	\end{figure}

\begin{figure}[h]
		\centering
		\includegraphics[width=0.45\textwidth]{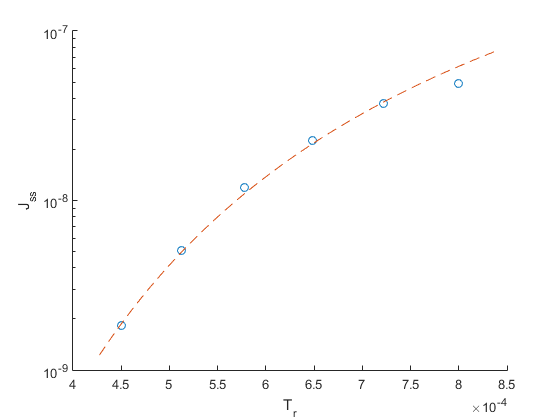}
		\caption{Plot of $J_{ss}$ for data sets $\kappa=0$ to $\kappa=5$ (blue circles), along with a fit of form $A_1\exp(-A_2/T_r)$ for fit parameters $A_1=5.57\times 10^{-6}$ and $A_2=0.0036$ (red dashed line). Error bars are not visible at this scale. }\label{fig:res_Jss_newnoise}
	\end{figure}

\begin{figure}[h]
		\centering
		\includegraphics[width=0.45\textwidth]{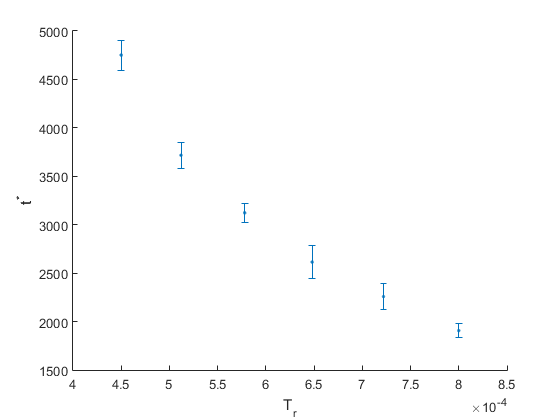}
		\caption{Plot of $t^*$ for data sets $\kappa=0$ to $\kappa=5$. Error bars were obtained from the uncertainty on the linear fit to the asymptote.}\label{fig:res_tinc_newnoise}
	\end{figure}

	We repeat the procedure of subsection~\ref{ssec:DB_varies} for the new data sets. Figure \ref{fig:res_I_datasets_newnoise} plots the post-critical nuclei densities $I^*(t)$ for each data set. Figures \ref{fig:res_Jss_newnoise} and \ref{fig:res_tinc_newnoise} plot $J_{ss}$ and $t^*$ for these data sets, respectively. Note that the x-axes for these two plots are in terms of $T_r=N_a^2/2$, the fluctuation temperature that enters the fluctuation-dissipation theorem as mentioned in subsection~\ref{ssec:time_pde}.

	Considering only the exponential terms of equation \ref{eq:nuc_Jss_scaling}, and recalling that $\gamma$ and $\Delta G$ only vary with $\Delta B$, the observed variation of $J_{ss}$ is expected to be due to $\Delta G_A$ and the dimensionless fluctuation temperature of the PFC model, $T_r$. We attempt a fit of form $A_1\exp(-A_2/T_r)$ to the plot of $J_{ss}$, where $A_1$ and $A_2$ are fit parameters, as shown in figure \ref{fig:res_Jss_newnoise}. The fit suggests agreement between equation \ref{eq:nuc_Jss_scaling} and our results, assuming that the factors of $1/T_r$ in the exponents are the main contributors to the variation of $J_{ss}$ over the considered range of $T_r$. The decrease of $\Delta G_A$ with increasing $T_r$ also possibly contributes to the observed increase in $J_{ss}$, though the effect of $1/T_r$ appears more prominently in the plot.
	
	As for incubation time, again due to $\gamma$ and $\Delta G$ not varying for these data sets, equation \ref{eq:nuc_tinc_scaling} indicates that $t^*$ would scale as $\exp(+\Delta G_A/k_B T)$ with $k_B T$ replaced by $T_r$ and $\Delta G_A$ decreasing with increasing $T_r$. Figure \ref{fig:res_tinc_newnoise} shows a decrease in $t^*$ as $T_r$ increases, which suggests qualitative agreement with equation \ref{eq:nuc_tinc_scaling}, though a specific fit was not attempted as the error bars allow both exponential as well as linear fits to be plausible.

	\subsection{Appearance and growth of lattice structure in early grains\label{ssec:structureform}}
	
	In CNT, the stochastically appearing grains are assumed to form with the same lattice structure as the final solid phase. Similarly, in the PFC model, the standard one-mode approximation for the solid phase also assumes the existence of well-defined lattice structure, as the PFC density $n(\vec{x})$ in the solid bulk can be expanded in terms of equal-amplitude Fourier wave modes corresponding to the lowest order reciprocal lattice vectors of the crystal structure (three wave modes for the case of a triangular lattice). However, it is feasible that free-standing planar waves or other non-lattice structures might temporarily appear in simulated PFC systems, especially during the early formation stages of solid grains (recall that amorphous structure has already been observed by other authors in both PFC model simulations \cite{toth10} as well as MD simulations \cite{ozgen04,tian08}). It is thus unclear whether the amplitude of all the one-mode approximation's wave modes must fluctuate to a nonzero value simultaneously and symmetrically for a stable solid grain to form from a liquid phase, or whether the wave modes can separately appear and build up to a stable nucleus over time. To better understand this early stage of grain formation in terms of the wave modes, and also to assess whether non-lattice structures were prevalent in the simulation runs used to obtain the data sets the previous subsections, we use a field filtering method to examine the growth of separate wave modes during the early formation of a few grains in PFC simulations.
	
	The field filtering method used in this subsection expands on work by Singer and Singer \cite{singer06}, where a method is developed to visualize the orientation of crystal grains in a fully solidified system. A `test wavelet' is constructed, with a density field given by a one-mode approximation corresponding to one of the solid phase's wave modes with a wave vector $\vec{q}$, multiplied to a Gaussian envelope. We convolve the test wavelet with the density field $n$ obtained from a simulation run, at a specific time $t$. This convolution enhances features of $n$ that exhibit the same structure as the wavelet. We then also apply a local averaging filter to smooth the wavelet-convolved $n$ field. The resulting filtered field's value provides at each spatial location a relative estimate of the amplitude of the wave mode corresponding to the wavelet's $\vec{q}$. By rotating the wavelet before applying this filtering process, the presence of a different wave mode can be examined at any location.
	
\begin{figure}[h]
		\centering
		\includegraphics[width=0.45\textwidth]{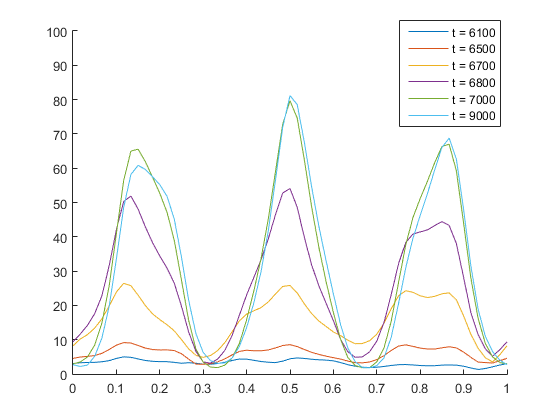}
		\caption{Value of the filtered field (y-axis) at the location of the first detected density peak of a forming grain, as a function of wavelet rotation angle (x-axis), for a range of times beginning before nucleation occurs and ending after the post-critical nucleus has grown to a size much larger than the critical size. The x-axis is in fractions of $\pi$. Time is in units of $d t$.}\label{fig:res_wavelet_fieldvalues}
	\end{figure}

	This filtering process is applied every few time steps of a nucleation-rate simulation, for a range of rotation angles, to obtain the relative amplitude of wave modes in the system as a function of time and position. We then store the values of the filtered fields at positions where nucleation occurs, taken to be the location of the first detected PFC density peak of each grain. Figure \ref{fig:res_wavelet_fieldvalues} plots the value of the filtered field at a specific location where a nucleation event was seen to occur, for a range of times and wavelet rotation angles, in a PFC system with model parameters corresponding to data set $\epsilon=4$ (see table \ref{tab:params_epsilon} for the parameters). The three peaks that emerge correspond to the three wave modes expected in the final solid bulk with triangular lattice structure (note that these displayed mode peaks should not be confused with the PFC density peaks that represent the position of atoms in the system). At the final time shown ($t=9000$), the post-critical nucleus is known to have grown to a size much larger than the critical size, indicating that the values of the filtered field at that time are that of the final solid. We note that the height of the three peaks in the final solid are not equal as would be expected from the standard one-mode approximation for the PFC model. This is assumed to be due to numerical error, as the square numerical grid has 4-fold symmetry while the final lattice structure has 6-fold symmetry, leading to slight numerical anisotropy in the application of the convolution filter.
	
\begin{figure*}[!ht]
		\centering
		\subfloat[]
		{
			\includegraphics[width=0.45\textwidth]{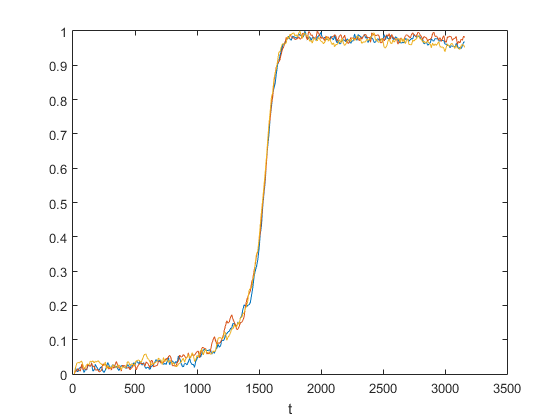}
		}\label{fig:res_convfield_nuc_a}
		\subfloat[]
		{
			\includegraphics[width=0.45\textwidth]{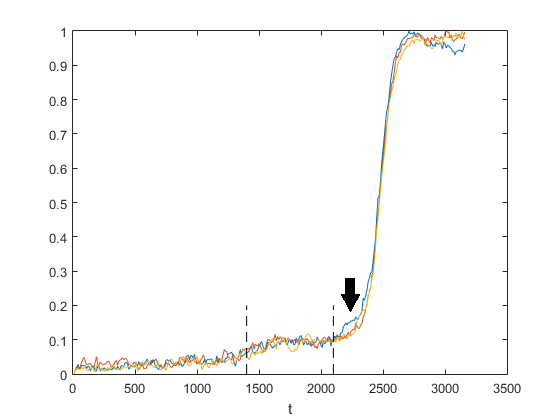}\label{fig:res_convfield_nuc_b}
		}
		\caption{Normalized values of the filtered field (y-axis) at the angles corresponding to the the peaks in the filtered field associated with a nucleation event  (corresponding to the three colours shown: red, yellow, and blue), as a function of time (in units of $d t$), for two separate nucleation events in the same simulated system.}\label{fig:res_convfield_nuc}
	\end{figure*}

	Once the angles for the peaks of the filtered field of a nucleation event are known at late times, we can plot the growth versus time of the field for only these three angles starting from early times. Figure \ref{fig:res_convfield_nuc} plots the growth of these peaks for two nucleation events, with parameters corresponding to data set $\epsilon=0$ in table \ref{tab:params_epsilon}. The values are normalized with respect to the maximum value attained by each peak, to account for the mentioned inequality of the peaks at late time due to numerical anisotropy. We also examined other nucleation events and obtained similar plots to these two cases.

	\begin{figure}[!h]
		\centering
		\includegraphics[width=0.45\textwidth]{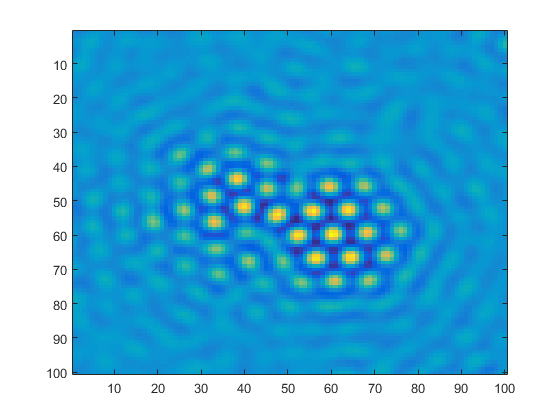}
		\caption{Density field of the grain from which  filtered field peak values shown in figure \ref{fig:res_convfield_nuc_b} where obtained, at simulation time $t=2550$. Axes indicate grid points.}\label{fig:res_convfield_examplegrain}
	\end{figure}

	\begin{figure}[h]
		\centering
		\includegraphics[width=0.45\textwidth]{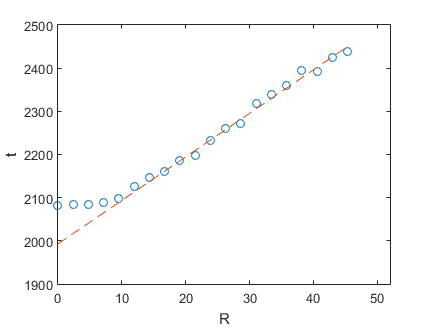}
		\caption{Simulation time taken for the density field of a forming grain to achieve half the maximum time correlation of the late-time (post full solidification) density field of that grain. The x-axis is radius in units of $dx$. The y-axis is time in units of $dt$. Blue circles are radial averages of the time taken, and red line is a linear fit to the large-radius averages.}\label{fig:res_timecorr1}
	\end{figure}
	
	The filtered field values at the three peak positions appear to vary in tandem during the majority of the process (up to the order of thermal fluctuations). This leads us to conclude that, at least for the parameter ranges of the data sets of section \ref{ssec:DB_varies}, nucleation in the PFC model exhibits a triangular atomic lattice structure even during the relatively early parts of grain formation. However, a minority of examined grains display unexpected behavior of the filtered field values at the three peaks, such as the grain corresponding to figure \ref{fig:res_convfield_nuc_b}. In that figure, the black arrow points to a relatively large fluctuation of only one mode that appears to precede the rapid growth of all three modes. The dashed lines denote a range of time where the three modes' growth seems to be delayed at a value higher than the liquid background value, yet lower than the final solid value. We believe these behaviors are due to a few grains forming with more complex forms, unaccounted for in our assumptions. Figure \ref{fig:res_convfield_examplegrain} shows the grain corresponding to figure \ref{fig:res_convfield_nuc_b}. We observe that this grain appears to exhibit two separate lattice orientations. This is possibly due to it being formed from two grains that merged into one at an earlier time. Another possible explanation is the existence of a precursor non-crystalline phase or preferred structure that precedes the critical nucleus. The competition between these separate lattice orientations might explain the growth delay observed in figure \ref{fig:res_convfield_nuc_b}, as well as the single mode fluctuation before the final rapid growth.
	
	These results indicate that the developed wave mode analysis method requires further refinement to be able to distinguish such edge cases. They also offer more insight into the difficulties involved in applying CNT to nucleation in the PFC model, as these complex-structured grains violate CNT's no-interaction or crystalline-structure assumptions, likely extending the required time for these grains to achieve criticality as their lattice structures stabilize.
	
	While the method above characterized the appearance of structure in grains by examining the growth of modes at only the first PFC density peak appearing in grains, we also briefly studied the spatial dependence of structure formation. By calculating the time correlation of the late-time (post full solidification) density field $n$ with earlier time values of the field, we obtain the relative rate of formation of lattice structure at spatial grid points at and near the forming grain. Figure \ref{fig:res_timecorr1} plots the average simulation time taken for the PFC density in a grain to achieve half its maximum time correlation with its late-time value, as a function of radial distance from the approximate center of the grain. The chosen grain was nearly circular, to ensure validity of radial averaging. We observe that, below a radius of approximately $10\,dx$ (corresponding to $1.25a$), the grain's structure approaches that of the final solid at a time independent of radius. Above this radius, information of the final lattice structure spreads linearly with time, as would be expected of a post-critical grain growing with constant interface velocity. The radius where this crossover behaviour occurs can be argued to be equivalent to the critical nucleus radius for the PFC model for the particular set of parameters used. This result provides further evidence that the appearance of a post-critical grain in the PFC model is preceded by fluctuation-induced lattice structure instantaneously appearing over a finite simulation volume. Though this method could in principle be repeated for a large number of circular grains at multiple simulation parameters to obtain statistics and parameter dependence for the critical radius, we do not attempt to do so in this work due to the difficulty in obtaining large numbers of sufficiently circular grains.
	
	\subsection{Numerical approximation for the form of critical nuclei}\label{ssec:critnuccurve}
	
	As discussed in subsection \ref{ssec:nuc_pfc}, we expect that the form of critical nuclei in the PFC model does not depend on only the number of atoms in an emerging grain, as assumed by CNT. We thus develop a method to efficiently numerically approximate the form of critical nuclei in the PFC model under the slightly more flexible assumption that both size and order of a grain can vary during the nucleation process. Note that Toth et al.~\cite{toth10} have previously examined the work of formation of critical nuclei by solving the Euler-Lagrange equation of the PFC model to obtain local extrema of the free energy functional. However, in this work we instead obtain approximate forms of the critical nuclei by numerically testing whether a constructed `test grain' of a given form is stable in a system with no fluctuations. While our method does not allow the reconstruction of the nucleation energy barrier, it provides insight on possible kinetics paths for a grain to attain criticality.
	
\begin{figure*}[!ht]
		\centering
		\subfloat[]
		{
			\includegraphics[width=0.45\textwidth]{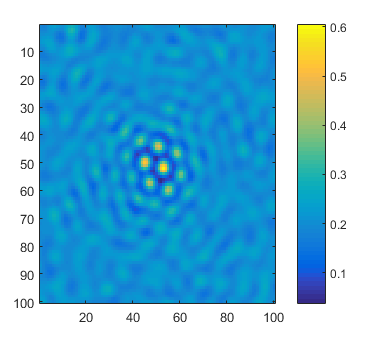}
		}
		\subfloat[]
		{
			\includegraphics[width=0.45\textwidth]{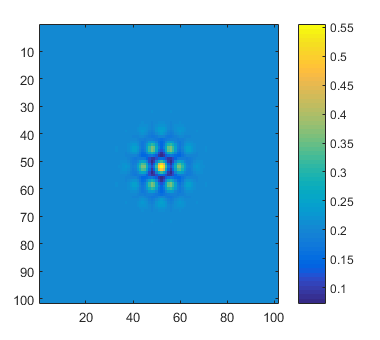}
		}
		\caption{(a) Solid grain surrounded by fluctuating liquid, observed in a simulation run with PFC model parameters $n_o=0.207$, $B^x=0.4$, $\Delta B=0.1650$, and $N_a=0.04$. (b) Constructed grain with same model parameters as (a) (without fluctuations), and with chosen parameters $r_1=0.5$ and $r_2=0.8$. In both figures, the x-axis and y-axis are in number of grid points, and the color bar shows the value of the density fields.}\label{fig:num_compareTestWavelet}
	\end{figure*}

	We start by constructing a `test grain', whose density field follows the lattice structure of a bulk PFC solid in the one-mode approximation, multiplied to a Gaussian envelope centered on one of the peaks. The amplitude of the test grain's density field is set to $r_1 \phi_s$, where $\phi_s$ is the amplitude predicted in a stable solid bulk at the system's position on its phase diagram, and $r_1$ is a value between 0 and 1. Similarly, the standard deviation of the Gaussian envelope is set to $r_2 a$ where $a$ is the dimensionless lattice constant and $r_2>0$. Figure \ref{fig:num_compareTestWavelet} shows a grain that formed stochastically in a simulated PFC run, as well as a comparable test grain constructed as explained above.
	
	Effectively, the parameter $r_1$ sets the relative order of the grain with respect to the original liquid and final solid phases, while $r_2$ determines the size of the grain. The parameter $r_1$ can also be heuristically understood as determining the average number of vacancies in the lattice of the forming grain, as it has been argued \cite{berry14} that variations in the amplitude of the PFC model's periodic density field can represent vacancy diffusion on diffusive time scales. We note that this approximate grain construction is only expected to be valid for small-radius grains; large stable solid grains would have a constant amplitude throughout their bulk and a finite interface width, rather than an approximately Gaussian profile for the amplitude. Furthermore, this construction does not account for a variation in \textit{average density} of the grain's interior, as the constructed periodic field would average to $n_o$ over a large enough bulk.
	
	For a given set of PFC model parameters and chosen $r_1$ and $r_2$, we can test whether a constructed test grain is post-critical or pre-critical by simulating it in a system consisting of the single grain in a much larger amount of liquid phase. The fluctuation amplitude is set to zero in this simulation, as the grain is assumed to be the result of a prior fluctuation, and we are interested in the subsequent deterministic evolution of this grain. If after sufficient time steps the grain has grown to fill the system with solid, then it is known to be a post-critical grain, and vice versa. By repeating this test for various $r_1$ and $r_2$ (using a half-interval search method in one of the two variables, for efficiency), we obtain a curve in the space spanned by these two parameters that determines whether a grain with specific $r_1$ and $r_2$ is post-critical. We refer here to this curve as `critical nucleus curve'. The exact form of the critical nucleus is thus not unique, as any set of $r_1$ and $r_2$ lying on such a curve would give a critical nucleus under the assumptions of the construction used.
	
\begin{figure}[h]
		\centering
		\includegraphics[width=0.45\textwidth]{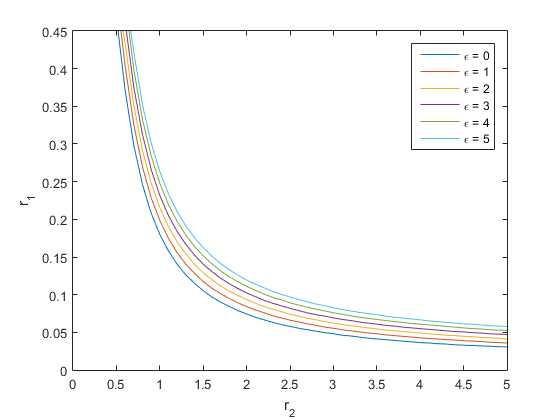}
		\caption{Critical nucleus curves for data sets $\epsilon=0$ to $\epsilon=5$. Grains with values of $(r_1,r_2)$ above the corresponding system's curve are post-critical. Recall that $r_1$ is a ratio that scales the amplitude of the periodic density field of the grain, while $r_2$ scales the radius of the Gaussian-shaped grain.}\label{fig:res_criticality}
	\end{figure}

	We numerically calculate the critical nucleus curves for the parameters of data sets $\epsilon=0$ to $\epsilon=5$ (see table \ref{tab:params_epsilon} for the parameters). Figure \ref{fig:res_criticality} shows these curves. Note that data sets $\kappa=0$ to $\kappa=5$ (see table \ref{tab:params_kappa} for the parameters) would have the same critical nucleus curve as data set $\epsilon=0$, as these curves do not depend on fluctuation amplitude $N_a$. We observe that, as $\epsilon$ (equivalently, the effective temperature $\Delta B$) increases, the curves shift along both the relative order axis (y-axis) and size axis (x-axis). The shift along the size axis agrees with the basic prediction of CNT that critical radius must increase with temperature. However, CNT does not account for the shift along the order axis, as it assumes the lattice structure in the interior of the grain is always the same as that of the final solid. Further, the precise kinetic path that a forming grain would take through $(r_1,r_2)$ parameter space before becoming a post-critical nucleus can not be predicted by CNT. This would instead require at least a 2 parameter theory similar to that developed in Ref.~\cite{lutsko15} for the case of nucleation in globular protein systems. We expect that the most likely path to criticality will depend on a balance between statistically probable fluctuation amplitude and diminishing spatial correlation at long range. Specifically, a critical nucleus of too small radius is unlikely to form due to the exponentially decreasing odds of obtaining a sufficiently large fluctuation as the required field amplitude increases, while a critical nucleus of too large radius is unlikely to form before smaller grains because the spatially conserved density fluctuations in the system limit the rate at which mass and information of the forming lattice structure can propagate at large distances. This appears to be supported by the data in Figure \ref{fig:res_timecorr1}, which suggests a  magnitude for this correlation length, and hence $r_2$.
%%%%%

	%%%%%%%%%%%%%%%%%%%%%%%%%%%%%%%%%%%%%%%%%%%%%%%%%%%%%%%%%%%%%%%%%%%%%%%%%%%%%%%%%%%%
	
	\section{Conclusion\label{sec:con}}
	
	The  goals of this work were to examine  time-dependent nucleation statistics in the PFC model and attempt a comparison with the predictions of CNT. We have shown that the PFC model follows the qualitative predictions of CNT. For fluctuations parameterized with model temperature such as to ensure correct capillary fluctuations \cite{kocher16}, homogeneous nucleation was found to  occur at strong undercoolings. The rate of nucleation was observed to not be constant, instead requiring a transient time to achieve its steady state behavior. The steady state rate of nucleation was shown to decrease at least exponentially as temperature increased, while incubation time was shown to increase nearly linearly with temperature (although no actual fit was attempted). All these behaviors were also argued to be qualitatively consistent with experimental nucleation rate predictions and results, within the constraint of negligible temperature dependence of mobility. Quantitative agreement with experiments would require significant tuning of the model that is beyond the scope of this work. Our results indicate that the PFC model can  be used to study solidification  phenomena that might require prior nucleation to initialize grain number density, such as structure growth, grain coalescence, and Ostwald ripening.
	
	Our results also showcased the CNT-predicted dependence of the steady state nucleation rate and incubation time explicitly on fluctuation amplitude. Despite the PFC model not including a direct equivalent of the CNT-assumed activation energy for atoms jumping through phase interfaces, the steady state nucleation rate was shown to vary with fluctuation amplitude following a dependence agreeing with that predicted for a thermally activated process. Similarly, the incubation time was seen to decrease as fluctuation amplitude increased, as expected in a system where propagation of mass and information is limited by the amplitude of spatially conserved density fluctuations.
	
	Finally, we also studied some of the limitations of CNT as applied to the PFC model. We examined the wave mode amplitudes in pre-critical grains, observing that, despite lattice structure appearing early on in the process for most grains, a minority of nucleation events displayed more complicated structural formation behavior that might affect the validity of growth rate and non-interaction assumptions used in CNT. We also numerically calculated `critical nucleus curves' to examine the approximate form of critical nuclei in the PFC model, under the assumption that both size and order of a grain are allowed to vary. These curves indicated that the CNT assumption of a single-parameter dependence (nucleus size) is likely insufficient to consistently predict nucleation in PFC. We suggest that a multi-parameter theory should be attempted, similar to the work in Ref.~\cite{lutsko15}. In the case of the PFC model, the parameters required might include some or all of the following: grain size, relative order compared to final solid state, local average density, and interface width.
	
	\bibliography{ref}
	
\end{document}